\documentclass[conference]{IEEEtran}
\IEEEoverridecommandlockouts
\usepackage{cite}
\usepackage{amsmath,amssymb,amsfonts}
\usepackage{algorithmic}
\usepackage{graphicx}
\usepackage{textcomp}
\usepackage{xcolor}
\usepackage{url}
\usepackage{hyperref}
\def\BibTeX{{\rm B\kern-.05em{\sc i\kern-.025em b}\kern-.08em
    T\kern-.1667em\lower.7ex\hbox{E}\kern-.125emX}}
\begin{document}

\title{Efficient Constraining of Transcoding in DNA-Based Image Storage
\thanks{This work is funded by the French ANR program ANR-22-PEXM-0005
MoleculArXiv.}
}
 
\author{\IEEEauthorblockN{Sara Al Sayyed}
\IEEEauthorblockA{Université de Rennes \\INRIA, France \\sara.al-sayyed@inria.fr}
\and
\IEEEauthorblockN{Aline Roumy}
\IEEEauthorblockA{Université de Rennes \\INRIA, France \\aline.roumy@inria.fr}
\and
\IEEEauthorblockN{Thomas Maugey}
\IEEEauthorblockA{Université de Rennes \\INRIA, France \\thomas.maugey@inria.fr}}


\def\ar#1{\textcolor{blue}{\footnotesize {(ar: #1)}}}

\maketitle
\begin{abstract}
DNA has emerged as a promising alternative for long-term data storage due to its high capacity, durability, and low-energy potential. However, storing data in DNA presents several challenges. First, it requires complex and costly biochemical processes, making efficient compression crucial to reducing DNA synthesis time and cost. Second, these processes are prone to errors that must be avoided and/or corrected. In particular, homopolymers (repetitions of the same nucleotide) are a well-known source of errors during the sequencing step. Avoiding such repetitions helps mitigate errors but introduces a constraint that may increase the data compression rate.
In this paper, we propose two transcoding methods that address these two key challenges: reducing data rate and minimizing errors. The first method strictly enforces the error-minimization constraint by eliminating homopolymers of a certain length, at the cost of an increased data rate. In contrast, the second method accepts a slight increase in homopolymers. However, we show that these increases remain limited (2.14\% increase in compression rate for the first method and 0.39\% homopolymer rate for the second). These two approaches demonstrate that it is possible to efficiently constrain transcoding while balancing error minimization and compression performance.
\end{abstract}


\section{Introduction}
The ever-growing use of data demands an ever-increasing number of servers, posing a significant environmental challenge.
As a result, DNA presents itself as a promising solution for long-term data storage because of its durability, high information density, and low-energy requirement \cite{b1}.
The typical process for DNA storage systems involves several key steps. First, the input data is encoded into a quaternary format using the four DNA bases: A, C, G, and T. These encoded sequences, called oligos, are then synthesized in a laboratory and stored in capsules within a protective environment. When retrieval is needed, specialized devices known as sequencers are used to read the DNA molecules. Finally, the sequenced data is decoded to reconstruct the original information.
The biochemical processes involved in DNA data storage is susceptible to three main types of errors: insertion, deletion, and substitution. There are two main approaches to handling these errors. One approach is to correct them using error correction codes, as demonstrated in \cite{b15}\cite{b18}. The other is to prevent them by satisfying some biochemical constraints during the encoding process. Ideally, an encoding scheme should both minimize and correct these errors to ensure reliable data storage.

Constraining the encoding process to prevent these errors can be achieved in two ways: internally, by incorporating constraints within the compression process, or externally, by transcoding the bitstream after compression.
%

Regarding the first approach, several algorithms have been proposed within the JPEG standardization research group (JPEG DNA \cite{b7}). For example, the Paircode dictionary \cite{b8} can constrain the output of various code types—variable-rate \cite{b8}, fixed-length \cite{b9}, or variable-length \cite{b9b}. Another approach constrains the tree construction for variable-length Fano codes \cite{b10}, applicable to encoding either a JPEG-like category or an autoencoder's latent representation \cite{b10b}. Finally, rotating dictionaries have been proposed to enforce constraints between codewords \cite{b11}.
%
However, these constrained compression schemes have a drawback: if error correction codes are applied to the encoded sequence, it may lose its constrained properties. In contrast, the encoding methods proposed by Goldman and Blawat \cite{b3}\cite{b5} use transcoding algorithms that enforce these constraints externally. However, these approaches sacrifice rate-efficiency to produce a constrained stream. 

%
In this paper, we propose two schemes that demonstrate it is possible to achieve rate-efficiency with constrained transcoding, which is consistent with the recent theoretical result \cite{b19weindalembracingerror} indicating that constrained coding is as efficient as error correction (when homopolymers longer than 4 are avoided, as is common in practice). The proposed methods follow two distinct approaches: the first provides a variable transcoding rate while maintaining a fixed level of constraint safety, making it ideal for storing critical data requiring high reliability. In contrast, the second maintains a fixed transcoding rate with a variable level of safety, making it suitable for applications needing predictable synthesis and sequencing costs. Both solutions are evaluated on image and video codecs (JPEG \cite{b20JPEGrecommendation}, VVC \cite{b21VVCstandarization}).

Specifically, in Section \ref{sec:context}, we  define the important constraints in DNA storage systems with baseline image coders and finally to pose the research problem to be tackled. Afterward, in Sections \ref{sec:breaksequence} and \ref{sec:dictionary}, we elaborate the proposed methods and analyze their transcoding rate and safety level. Finally, in Section \ref{sec:results}, we show the performance results of the proposed methods applied for image and video coders.

\section{Constrained Coding on DNA: a trade-off between safety and efficiency?}
\label{sec:context}
In the process of DNA data storage, errors often arise due to specific structures and patterns within the DNA sequence that must be avoided. To mitigate this, certain constraints are applied during the coding process. Failure to adhere to these constraints can significantly increase the likelihood of errors. The primary constraints are as follows:  
\begin{itemize}
    \item \textbf{Homopolymers}: The consecutive occurrence of the same nucleotide should not exceed three repetitions.
    \item \textbf{Balanced GC content}: The proportion of guanine (G) and cytosine (C) in the sequence must fall between 40\% and 60\%.
\end{itemize}

In this context, the concept of safety can be defined as follows: \textit{an encoding algorithm is considered safe if it adheres to these constraints.}

Similarly, the notion of efficiency is defined as: \textit{an efficient encoding algorithm is one that achieves a low compression rate, measured in nucleotides per pixel.}

In our research, we have considered two baseline image coders that are both representative and exemplify extreme differences based on the above criteria:
\begin{itemize}
    \item \textbf{JPEG-DNA}: A JPEG-inspired codec \cite{b9} that ensures safety by integrating biochemical constraints directly within the image compression process.  
    \item \textbf{JPEG-DirectTranscoding}: This method involves encoding the image using JPEG in the first stage, followed by transcoding the output bitstream using a direct approach that maps each bit pair to one nucleotide without considering the constraints mentioned above. While efficient, this approach is not safe.
\end{itemize}


This leads to a key research question: \textit{does adhering to safety constraints mean sacrificing rate efficiency, or can we develop encoding methods that achieve a trade-off between efficiency and safety? } 

In the rest of this paper, we will focus on exploring achievable and non-complex transcoding methods that satisfy safety and efficiency. For these methods, we will use the \emph{transcoding rate metric}, measured in nucleotides per bit, which indicates the number of nucleotides required to represent one bit.

\textbf{Assumption i.i.d. uniform}: The transcoding methods are applied to already compressed sequences. We assume that the compression schemes are optimal, so the sequences are independent and uniformly distributed (i.i.d.). We verified this assumption on JPEG-encoded Kodak dataset\footnote{\url{http://r0k.us/graphics/kodak/}} bitstreams, which are nearly i.i.d. uniform.

\section{Variable Transcoding Rate - fixed Safety}
\label{sec:breaksequence}
\subsection{The Method: Break-Sequence}
In this section, we propose a method adaptable to the encoded stream, aiming to ensure perfect biological safety while optimizing efficiency. Specifically, we suggest breaking the encoded sequence and introducing an additional nucleotide only when necessary.

To clarify, after encoding the input (e.g., an image) using JPEG, we propose transcoding the resulting bitstream into a nucleotide stream using a break-sequence approach. In this method, each bit pair in the bitstream is transcoded using a direct rule (00$\to$A,01$\to$T,10$\to$C,11$\to$G) until $K-1$ consecutive identical nucleotides are encountered. At this point, a single bit (b) is transcoded into one nucleotide, say $n_i$.

This single-bit transcoding relies on the surrounding nucleotides ($n_{i-1}$ and $n_{i+1}$) and follows these rules:

\begin{itemize}
\item \textbf{If $n_{i-1}$ and $n_{i+1}$ are identical:} Transcode b using dictionary Dic1 in Table \ref{tab:bittrans}, selecting an option different from $n_{i-1}=n_{i+1}$ (i.e. choose O1 if $n_{i-1}\in$ O2).
\item \textbf{If $n_{i-1}$ and $n_{i+1}$ are different:} Select a dictionary from Table \ref{tab:bittrans} where $n_{i-1}$ and $n_{i+1}$ are in the same column, then choose an option different from both $n_{i-1}$ and $n_{i+1}$.
\end{itemize}
\emph{Example:} The sequence '00 00 00 1 00' is encoded as 'AAAGA', where the 'G' is generated by encoding the bit '1' using option O2 in Dic1. In contrast, the sequence '00 00 00 1 11' is encoded as 'AAATG'; here, the bit '1' is encoded using Dic3 option O2 because the pair $(n_{i-1},n_{i+1})$=(A,G) belongs to Dic3 option O1.

By applying this method, we ensure that no sequences containing more than three consecutive identical nucleotides are created, effectively preventing the formation of homopolymers.

This approach is perfectly safe, but the rate may become high if the threat of homopolymers (three consecutive identical nucleotides) occurs frequently. In the following, we propose a study that demonstrates that the rate overhead remains limited.

\begin{table}[htbp]
\caption{Bit Transcoding}
\label{tab:bittrans}
\centering
        \begin{tabular}{|c|cc|cc|cc|}
        \hline
            
           \textbf{}&\multicolumn{2}{|c|}{\textbf{Dic1}}&\multicolumn{2}{|c|}{\textbf{Dic2}}&\multicolumn{2}{|c|}{\textbf{Dic3}}
           \\
            \hline
            \textbf{Bit \textbackslash Option} & \textbf{O1} & \textbf{O2} & \textbf{O1} & \textbf{O2} & \textbf{O1} & \textbf{O2}\\
            \hline
            0 & A & C & A & T & A & C \\
            1 & T & G & C & G & G & T \\
            \hline
            
        \end{tabular}
\end{table}

\subsection{Analytical Study}
\textbf{Principle: } If $K-1$ consecutive nucleotides are equal, transcode the next bit into $n_{i}$ such that
$n_{i} \neq (n_{i-1}, n_{i+1})$
\subsubsection{Transcoding Rate}
The transcoding rate can be expressed as follow:
\begin{equation}
    R=(1-\varepsilon) \times \frac{1}{2}+\varepsilon \times 1 
\end{equation}
where \( \varepsilon\) is the probability of having K-1 consecutive nucleotides identical (in our case K=4)
\begin{equation}
  \begin{aligned}
     \varepsilon &= P (\text{ K-1 consecutive identical nts } )\\
    &= \sum_{x \in \{A, C, G, T\}} P\big(X_{i-K+1} = x, \ldots, X_{i-1} = x\big)\\
    &=  {\left(\frac{1}{4}\right)}^{K-2}    \label{eq2}
\end{aligned}  
\end{equation}
where the second equation in \eqref{eq2} follows from the i.i.d. uniform assumption.
Taking K=4, the value of R=0.53125, which is slightly higher than the transcoding rate of the direct transcoding method (0.5) having maximum efficiency. This suggests that the rate overhead is small, making this approach highly efficient.
Additionally, we calculated the Bjøntegaard metric (BD-Rate) between the theoretical and experimental RD-curves for kodim23 image, and the results showed a BD-Rate of 3.897\%. This shows that our estimation for the transcoding rate approaches the real one. 

\subsubsection{Percentage of nucleotides forming homopolymer}
Clearly, we ensure that no more than three identical nucleotides occur consecutively in this method. Therefore:
\[
P_{h}=0
\]
\label{subsec:homobreak}

\section{Fixed Transcoding Rate - Variable Safety}
\label{sec:dictionary}
\subsection{The Method: Tuned-Dictionary}
Aiming to fix the rate overhead upon transcoding, while being more safe in terms of constraints when compared to direct transcoding, we propose the tuned dictionary-based approach. In this method, we introduce the use of an additional nucleotide for a configurable number of bits, denoted as \( P \). This parameter \( P \) can be adjusted based on specific requirements.

Using the four dictionaries detailed in Table~\ref{tab:dictionaries}, each group of \( P \) bits is transcoded into \( \frac{P}{2} + 1 \) nucleotides, where the extra nucleotide indicates the dictionary used for transcoding. These dictionaries are designed to enhance the diversity of the resulting sequences. Specifically, the first dictionary maps each bit pair to a single, fixed nucleotide, while the remaining three dictionaries provide four possible nucleotide options for each bit pair. The selection among these options is determined by the position of the bit pair within the \( P \)-bit group.

To explain further, the input bitstream is divided into groups of \( P \) bits, where \( P \) is an even number. From each group, \( \frac{P}{2} \) bit pairs are formed. These bit pairs are then encoded iteratively using the four dictionaries. Initially, the first dictionary is used to map each pair to its corresponding nucleotide. For the other three dictionaries, the nucleotide is chosen based on the column corresponding to \( \text{pos(pair)} \mod 4 \), where \( \text{pos(pair)} \) is the position of the pair within the group.

After encoding the \( P \) bits using all four dictionaries, we select the encoding with the smallest average homopolymer length. If multiple encodings result in the same minimum average homopolymer length, we choose the encoding that does not begin with a homopolymer, reducing the risk of forming longer homopolymers during concatenation.

\begin{table}[htbp]
    \centering
    \caption{Dictionaries for Transcoding}
    \label{tab:dictionaries}
    \setlength{\tabcolsep}{6pt} 
    \renewcommand{\arraystretch}{1.2} 

    \begin{minipage}{0.46\linewidth}
        \centering
        \begin{tabular}{c|cccc}
            
            \textbf{Pair} & \textbf{N0} & \textbf{N1} & \textbf{N2} & \textbf{N3} \\
            \hline
            00 & A & A & A & A \\
            01 & T & T & T & T \\
            10 & C & C & C & C \\
            11 & G & G & G & G \\
           
        \end{tabular}
    \end{minipage}
    \hfill
    \begin{minipage}{0.5\linewidth}
        \centering
        \begin{tabular}{c|cccc}
            
            \textbf{Pair} & \textbf{N0} & \textbf{N1} & \textbf{N2} & \textbf{N3} \\
            \hline
            00 & A & T & C & G \\
            01 & T & C & G & A \\
            10 & C & G & A & T \\
            11 & G & A & T & C \\
           
        \end{tabular}
    \end{minipage}

    \vspace{10pt}

    \begin{minipage}{0.46\linewidth}
        \centering
        \begin{tabular}{c|cccc}
            
            \textbf{Pair} & \textbf{N0} & \textbf{N1} & \textbf{N2} & \textbf{N3} \\
            \hline
            00 & A & G & T & C \\
            01 & C & A & G & T \\
            10 & G & T & C & A \\
            11 & T & C & A & G \\
            
        \end{tabular}
    \end{minipage}
    \hfill
    \begin{minipage}{0.5\linewidth}
        \centering
        \begin{tabular}{c|cccc}
            
            \textbf{Pair} & \textbf{N0} & \textbf{N1} & \textbf{N2} & \textbf{N3} \\
            \hline
            00 & G & A & T & C \\
            01 & T & C & G & A \\
            10 & A & G & C & T \\
            11 & C & T & A & G \\
            
        \end{tabular}
    \end{minipage}
\end{table}

\subsection{Analytical Study}
\textbf{Principle:} Transcode \textbf{$P$} bits to \textbf{$P/2+1$} nucleotides using four dictionaries.
\subsubsection{Transcoding Rate}
\begin{figure}[htbp]
\centerline{\includegraphics[width=0.45\textwidth]{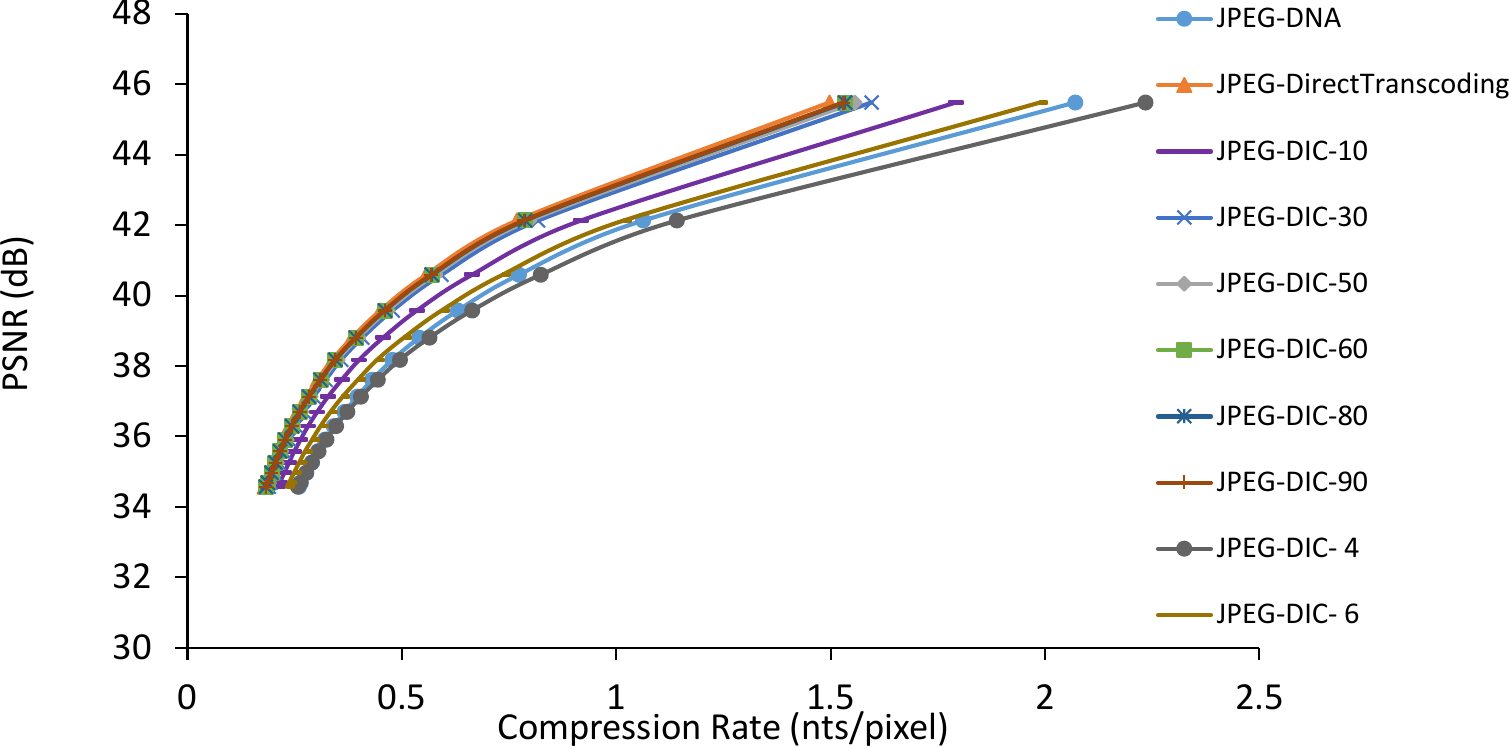}}
\caption{Compression performance comparison for different value of $P$ in the tuned-dictionary method for kodim23}
\label{fig1}
\end{figure}

The transcoding rate of this method can be expressed as follows:
\begin{equation}
R=\frac{\frac{P}{2}+1}{P}   \label{eq1}
\end{equation}
As \( P \to \infty \), \eqref{eq1} tends to 1/2, which corresponds to the transcoding rate of the direct method with maximum efficiency. This demonstrates that as $P$ increases, the efficiency of this method improves, approaching that of the direct transcoding method. This behavior is clearly shown in Figure \ref{fig1}.
\subsubsection{Percentage of nucleotides forming homopolymers}
\begin{figure}[htbp]
\centerline{\includegraphics[width=0.45\textwidth]{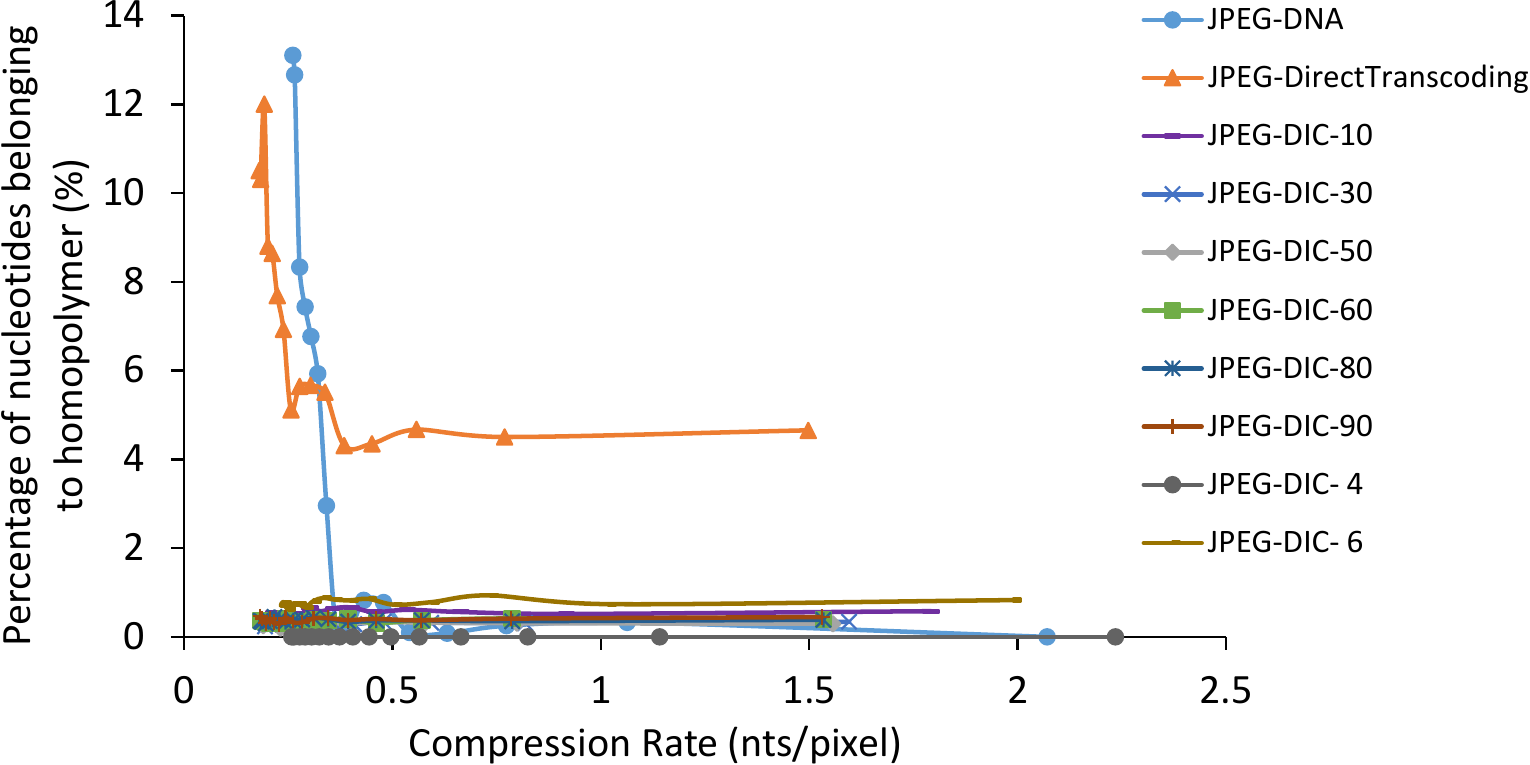}}
\caption{Homopolymer Safety comparison for different values of $P$ in the tuned-dictionary method for kodim23}
\label{fig2}
\end{figure}

In this method, we aim to minimize the formation of homopolymers by selecting sequences with the smallest average homopolymer length. However, homopolymers may still occur. As shown in Figure~\ref{fig2}, if \(\ P\) $>$ 4, the percentage of nucleotides forming homopolymers is not directly related to the value of \( P \). In the following, we aim to estimate the percentage of nucleotides that belong to homopolymers of length $K$ for $P > 4$.

Typically, we can express the expected number of homopolymers of length \( K \) in an i.i.d. uniformly distributed nucleotide sequence of length \( N \) as:

\begin{equation}
\label{eq4}
E(N_{HK}) = \frac{(N - K + 1) \times 4}{4^K} = \frac{N - K + 1}{4^{K-1}}
\end{equation}
where
\begin{itemize}
    \item \( N - K + 1 \) represents the number of possible homopolymers in a sequence of length \( N \).
    \item The probability of obtaining a homopolymer of length \( K \) is \( \frac{4}{4^K} \). The factor 4 accounts for the four possible homopolymers: AAAA, TTTT, CCCC, and GGGG.
\end{itemize}
Since this method employs four dictionaries, our initial approximation of the percentage of nucleotides that belong to homopolymers in the dictionary-based approach is:
\begin{equation}
\tilde P_{D} \approx \frac{E(N_{HK})\times K}{4\times N} \times 100 =\frac{(N - K + 1) \times K}{4^{K-1} \times 4 \times N} \times 100
\end{equation}
However, this approximation tends to be pessimistic since it doesn't take into account the construction of the dictionaries. Referring to the dictionaries in Table~\ref{tab:dictionaries}, we observe that for each bit pair at a given position, there are approximately two possible nucleotide options. For example, for a bit pair '00' at position \( b = 0 \), we can substitute it with two options: {A, G}. 
Thus, to capture the diversity of our proposed dictionary-based transcoding method, we approximate the encoding process as follows: each $N_{i}$ in the nucleotide stream can be seen as obtained from a bit pair and a transcoding function as follow:
\begin{equation}
\label{eq5}
    N_{i}=B_{2i}B_{2i+1}T_{i}
\end{equation}
where $B_{2i}B_{2i+1}$ is the original bitstream 
    and $T_{i}$ is the transcoding function. Due to the i.i.d. uniform assumption (see Section~\ref{sec:context})
    $B_{2i} \sim \text{Bernoulli}\left(\frac{1}{2}\right)$ and
    $B_{2i+1} \sim \text{Bernoulli}\left(\frac{1}{2}\right)$, and they are independent. Furthermore, the diversity of the dictionary is expressed as 
    $T_{i} \sim \text{Bernoulli}\left(\frac{1}{2}\right)$. Then, we disregard the fact that the dictionary selection depends on the bitstream and assume that transcoding is independent of bitstream formation, i.e., $T_{i}$ is independent of $B_{2i}B_{2i+1}$. Therefore,
\begin{equation}
\label{eq6}
P(N_{i} = N) = \frac{1}{2}*\frac{1}{2}*\frac{1}{2} = \frac{1}{8}
\end{equation}
Based on \eqref{eq4} and \eqref{eq6}, we can approximate the expected number of homopolymers of length \( K \) in a sequence of length \( N \) after applying the tuned-dictionary method as follows:

\begin{equation}
\label{eq7}
E(N_{HKD}) \approx \frac{(N - K + 1) \times 4}{8^K}
\end{equation}
Thus, we can conclude the approximate percentage of nucleotides forming homopolymers of length K as:
\begin{equation}
\label{eq8}
P_D \approx \left( \frac{(N - K + 1) \times 4\times K}{8^{K}  \times N} \right) \times 100
\end{equation}

\textbf{Accuracy}: We now wish to test the accuracy of our approximation in \eqref{eq7} and \eqref{eq8}. To do so, we experimentally evaluated the expected number of homopolymers and the percentage of nucleotides belonging to a homopolymer through 10,000 trials. In each trial, a uniformly distributed bit sequence was generated and transcoded using the tuned-dictionary method at $P=50$ to produce a nucleotide sequence of length $N=5000$. We set the homopolymer size to be avoided to $K=4$, as this is a typical value. Finally, we obtained the following results:
\begin{itemize}
    \item \textbf{Experimental values:} the expected number of homopolymers =4.3741, percentage=0.3248\%
    \item \textbf{Theoretical approximations} of \eqref{eq7} and \eqref{eq8}: the expected number of homopolymers=4.8798, percentage = 0.39039\%
\end{itemize}
We conclude that the theoretical approximations are very close to the true experimental values.



\begin{figure}[htbp]
\centerline{\includegraphics[width=0.45\textwidth]{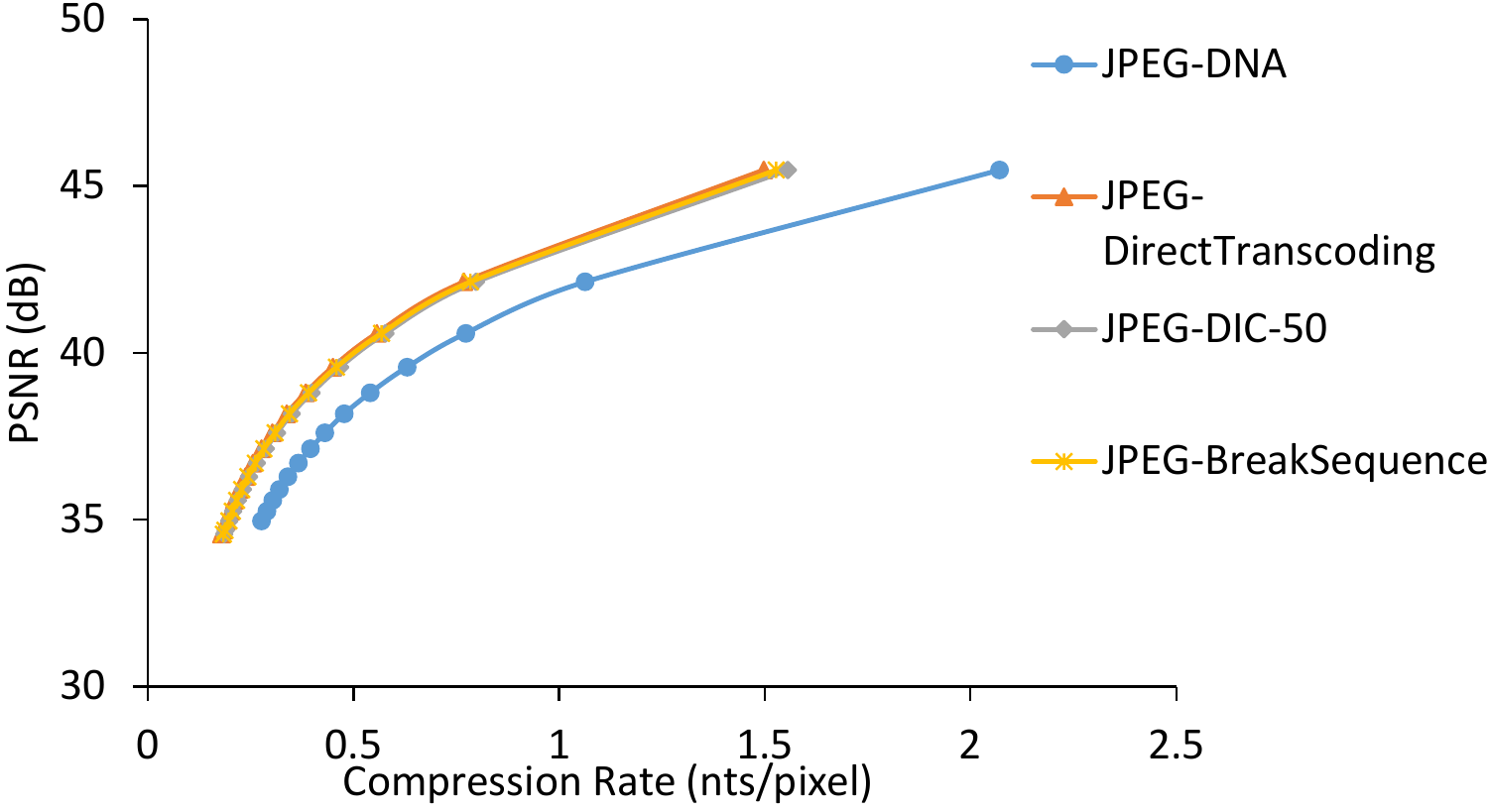}}
\caption{Compression Performance comparison for the image kodim23 from the kodak dataset.}
\label{fig3}
\end{figure}
\begin{figure}[htbp]
\centerline{\includegraphics[width=0.45\textwidth]{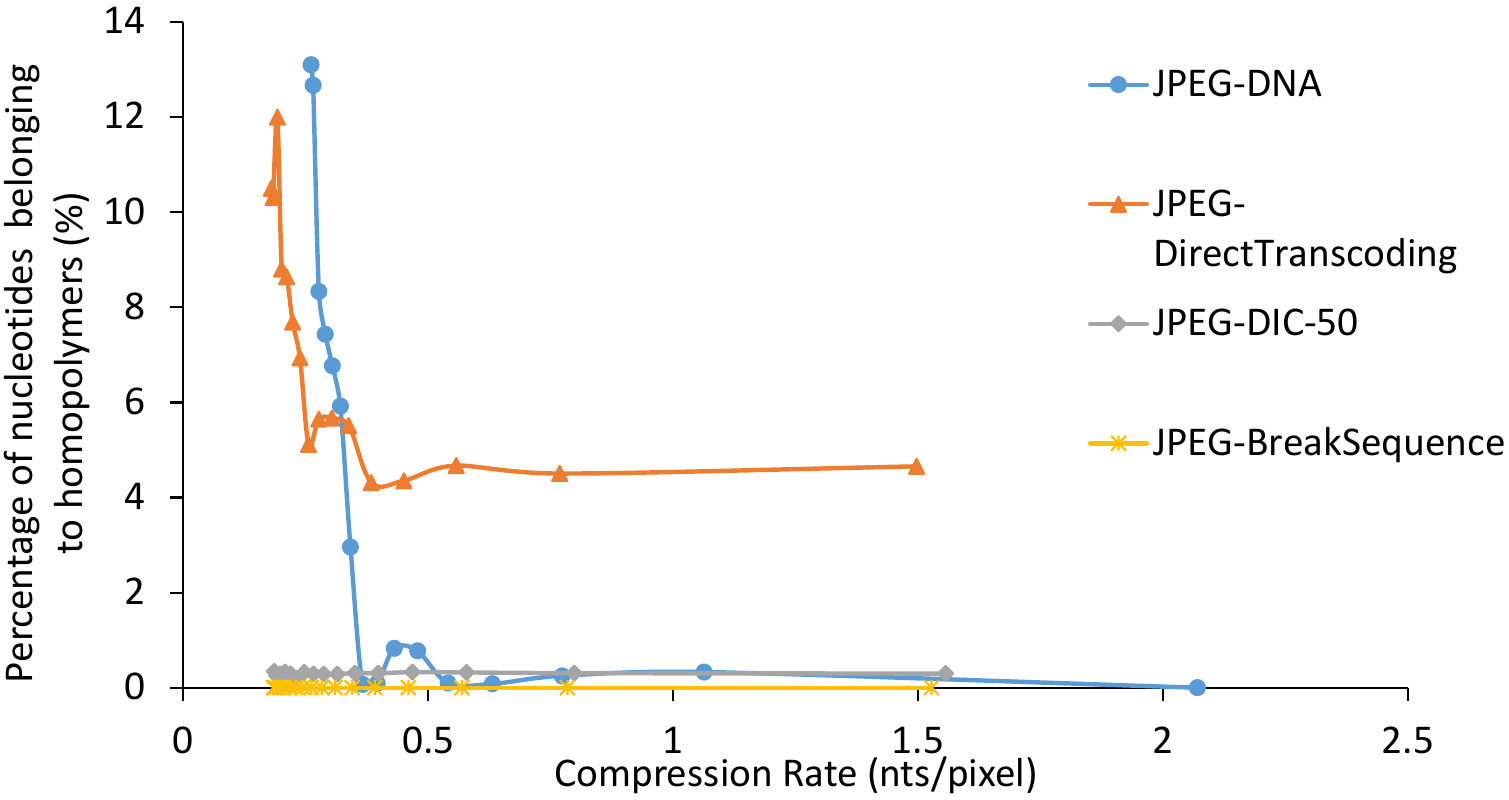}}
\caption{Homopolymer Percentage
vs. Compression Rate for kodim23: A Homopolymer Safety Comparison}
\label{fig4}
\end{figure}
\begin{figure}[htbp]
\centerline{\includegraphics[width=0.45\textwidth]{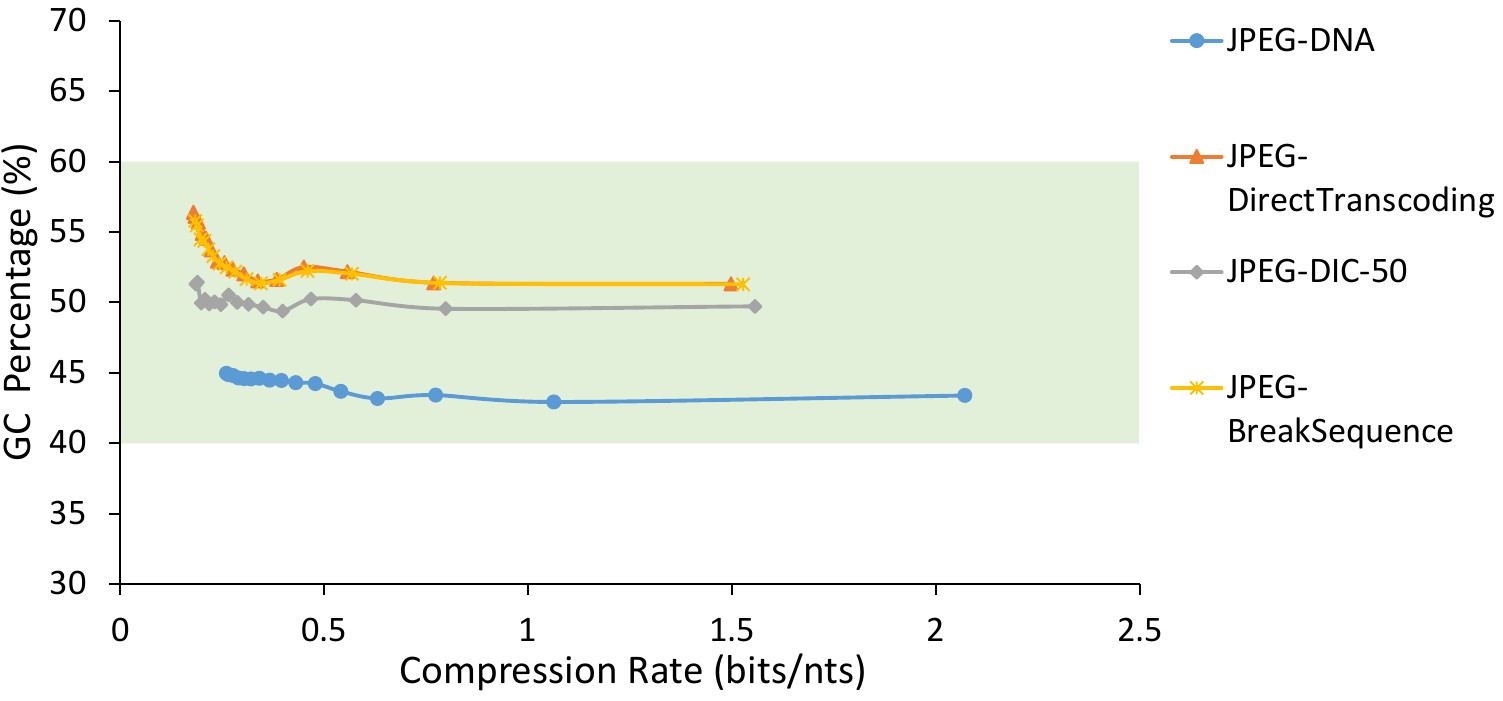}}
\caption{GC Percentage vs. Compression Rate for kodim23: A GC Safety Comparison}
\label{fig5}
\end{figure}
\begin{figure}[htbp]
\centerline{\includegraphics[width=0.45\textwidth]{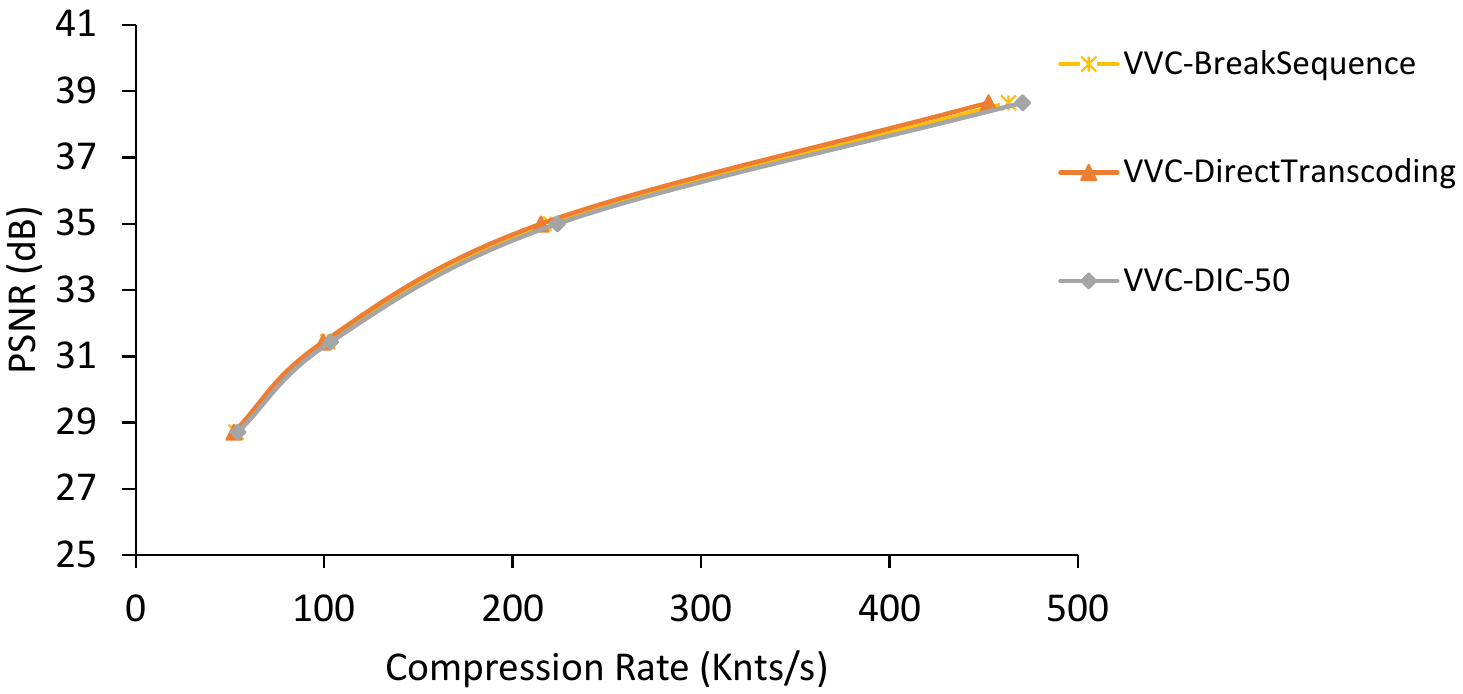}}
\caption{Compression Performance on the 'bus' video  from the Xiph dataset.}
\label{fig6}
\end{figure}
\begin{figure}[htbp]\centerline{\includegraphics[width=0.45\textwidth]{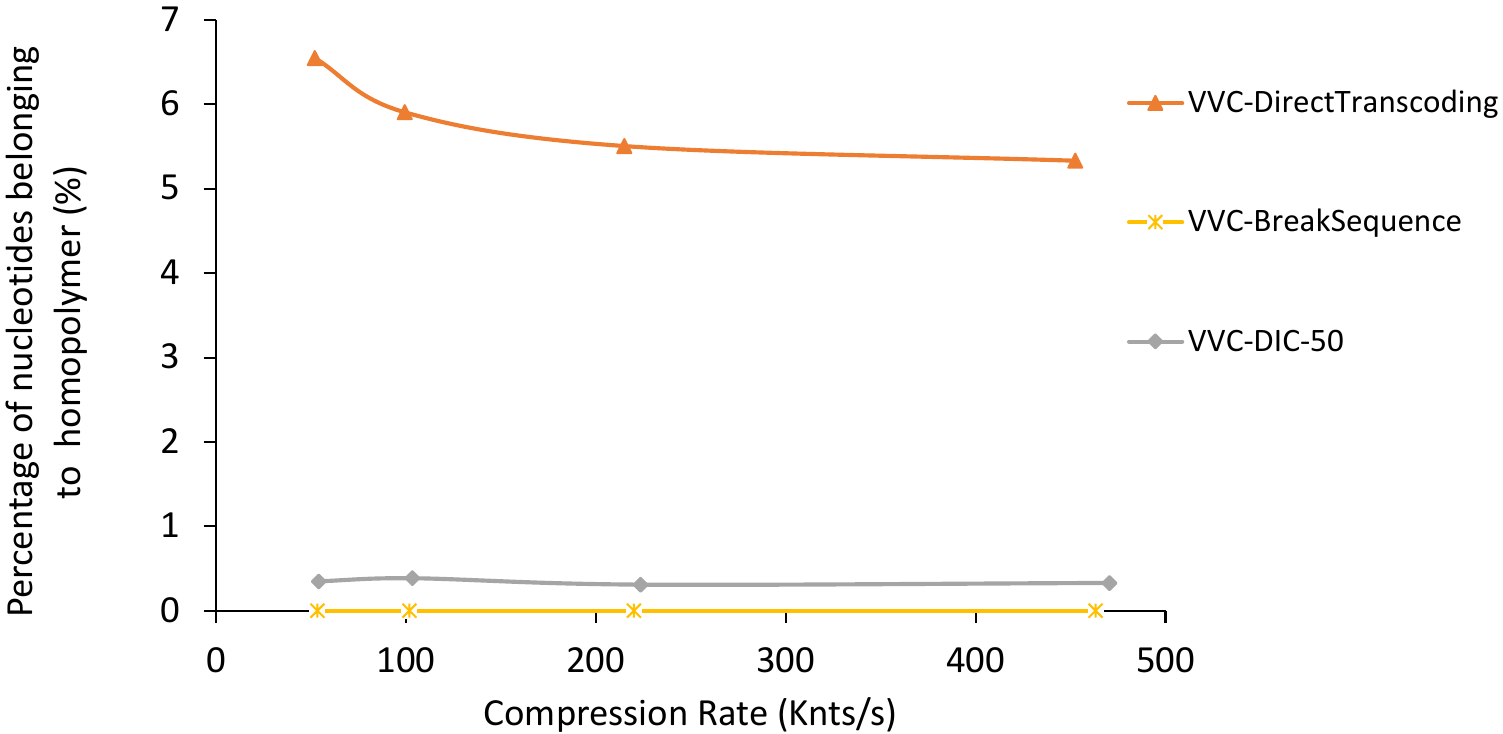}}
\caption{Homopolymer Percentage vs. Compression Rate for 'bus' video from the Xiph dataset.}
\label{fig7}
\end{figure}

\section{Performance Results}
\label{sec:results}
\subsection{Compression Performance}
We tested the performance of the proposed methods using the Kodak dataset and selected the dictionary method with $P = 50$. As shown in Figure~\ref{fig3}, applying the proposed transcoding methods after JPEG (JPEG-DIC-50 and JPEG-BreakSequence) demonstrates improved efficiency compared to JPEG-DNA, gradually approaching the efficiency of the direct transcoding method.
\subsection{Safety Performance}
\subsubsection{Homopolymers}
In Figure \ref{fig4}, we show the percentage of nucleotides forming homopolymers for the kodim23 image in the Kodak dataset, plotted against the compression rate in nts/pixels. It is evident that the proposed methods applied after JPEG offer a significant advantage in terms of homopolymer safety. Specifically, by using the dictionary method with $P = 50$, we achieve a reduced percentage of nucleotides forming homopolymers compared to JPEG-DNA at low compression rates. Moreover, the break-sequence method, as described in section \ref{sec:breaksequence}, achieves the best performance by eliminating this percentage entirely.

\subsubsection{GC Content}
In Figure \ref{fig5}, we plot the percentage of G and C nucleotides for the kodim23 image. It is clear that all the methods studied satisfy the GC percentage constraint mentioned in section \ref{sec:context}. These results were the primary motivation for designing transcoding methods that specifically address the issue of homopolymers.

\subsection{Performance on Videos}
We implemented the proposed transcoding methods on a VVC-compressed video (``Bus") consisting of 150 frames with a resolution of 352×288. As shown in Figure \ref{fig6}, the compression performance of our methods still very high, approaching that of direct transcoding. Additionally, from Figure \ref{fig7}, we observe that the percentage of nucleotides forming homopolymers for the dictionary method at P=50 is very low. Notably, this percentage remains consistently zero for the break-sequence method. These results verify that our methods achieve high levels of safety and efficiency, even when applied to large files such as videos.

\section{Conclusion}

In this paper, we studied the trade-off between rate efficiency and safety in DNA storage, where a scheme is considered safe if the produced nucleotide sequence satisfies biochemical constraints, resulting in a low probability of errors. We proposed two transcoding schemes: 
We theoretically and experimentally evaluated the excess rate of the first method and the probability of violating safety constraints for the second, demonstrating that both remained within acceptable limits. Moreover, the BreakSequence approach provided the best trade-off between rate efficiency and safety. This trade-off was assessed on JPEG-encoded images as well as VVC-encoded videos.


 Beyond the rate efficiency-safety trade-off, the proposed methods offer several additional advantages. First, they are transcoding-based, making them compatible with existing techniques like compression and error correction. Second, they are instantaneous, allowing decoding without waiting for the full nucleotide sequence. In the BreakSequence method, only one additional nucleotide needs to be read in advance before starting the decoding process. Third, they have low computational complexity, comparable to that of the direct transcoding method. 

This question of complexity is particularly important. A recent study found that avoiding homopolymers (length $>$ 4) and relying only on error correction achieve similar efficiency, but pure error correction was preferred due to its lower computational costs. This paper proposes a low-complexity constrained coding method, positioning it as a complementary approach rather than a competitor to error correction.

Finally, as future work, it would be interesting to integrate these methods with existing DNA storage systems. Additionally, testing the effect of homopolymers on the error rate of these systems would provide valuable insights. It would also be beneficial to evaluate how the proposed transcoding methods perform with different sequencing technologies.

\section*{Acknowledgment}
The authors would like to thank all the members of MoleculArXiv project, led by Marc Antonini. Special thanks also go to Dominique Lavenier  for his valuable inputs on this work.







\end{document}